\documentclass[%
preprint%
 ,secnumarabic%
,superscriptaddress%
,amssymb, amsmath,nobibnotes,aps]{revtex4}

\usepackage{epsfig}%
\usepackage{graphicx}%
\expandafter\ifx\csname package@font\endcsname\relax\else
 \expandafter\expandafter
 \expandafter\usepackage
 \expandafter\expandafter
 \expandafter{\csname package@font\endcsname}%
\fi

\begin{document}

\title{Constraints on background torsion from birefringence of CMB polarization}
\author{Moumita Das}
\email{moumita@prl.res.in}
\affiliation{Physical Research Laboratory, Ahmedabad 380009,
India}
\author{Subhendra Mohanty}
\email{mohanty@prl.res.in}
\affiliation{Physical Research Laboratory, Ahmedabad 380009,
India}
\author{A.R.Prasanna}
\email{prasanna@prl.res.in}
\affiliation{Physical Research Laboratory, Ahmedabad 380009,
India}
\affiliation{L.J.Institute of Computer Application, Ahmedabad, India}
\def\be{\begin{equation}}
\def\ee{\end{equation}}
\def\al{\alpha}
\def\bea{\begin{eqnarray}}
\def\eea{\end{eqnarray}}

%%%%%%%%%%%%%%%%%%%%%%%%%%%%%%%%%%%%%%%%%%%%%%%%%%%%%%%%%%%%%%%%%%%%%%%%%%%%%%%%%%%%%%%%%%%%%%%%%%%%%%

\begin{abstract}
We show that a non-minimal coupling of electromagnetism with
background torsion can produce birefringence of the electromagnetic
waves. This birefringence gives rise to a B-mode polarization of the
CMB. From the bounds on B-mode from WMAP and BOOMERanG data, one can
put limits on the background torsion at $\xi_{1}T_{1}=\left(-3.35 \pm
2.65\right)\,\times\,10^{-22}\,\,\,GeV^{-1}$.
\end{abstract}
\maketitle

%%%%%%%%%%%%%%%%%%%%%%%%%%%%%%%%%%%%%%%%%%%%%%%%%%%%%%%%%%%%%%%%%%%%%%%%%%%%%%%%%%%%%%%%%%%%%%%%%%%%%%
\section{Introduction}
In general relativity, the connection is defined through the covariant derivatives,
\begin{eqnarray}
\nabla_{\mu}V_{\nu}=\partial_{\mu}V_{\nu}+\Gamma^{\alpha}_{\,\,\,\mu\nu}V_{\alpha}
\end{eqnarray}
$\Gamma^{\alpha}_{\,\,\,\mu\nu}$ is symmetric in lower indices and is known as Levi-Civita connection.
On the other hand, in Einstein-Cartan theory \cite{Trautman}, which is an extension of general relativity, $\Gamma^{\alpha}_{\,\,\,\mu\nu}$
is asymmetric with its symmetric part $\left\lbrace{^{\alpha}_{\mu\nu}}\right\rbrace$, the Levi-Civita part, also 
known as the Christoffel symbol and the torsion part $ T^{\alpha}_{\,\,\,\mu\nu}$ which is antisymmetric in $\mu$
and $\nu$ as given by,
\begin{eqnarray}
\Gamma^{\alpha}_{\,\,\,\mu\nu}\equiv \left\lbrace{^{\alpha}_{\mu\nu}}\right\rbrace+T^{\alpha}_{\,\,\,\mu\nu}
\end{eqnarray}
and
\begin{eqnarray}
\left\lbrace{^{\alpha}_{\mu\nu}}\right\rbrace =\frac{1}{2} g^{\alpha \lambda}\left(\partial_{\mu}
g_{\lambda \nu}+\partial_{\nu} g_{\lambda \mu} - \partial_{\lambda} g_{\mu \nu} \right)\,\,,\,\,
T^{\alpha}_{\,\,\,\mu\nu}=\frac{1}{2}\left(\Gamma^{\alpha}_{\,\,\,\mu\nu}-\Gamma^{\alpha}_{\,\,\,\nu\mu}\right).
\end{eqnarray}
In this paper we shall consider the generalized connection restricting ourselves to metric theories such that we still have
$\nabla_\mu g_{\alpha \beta}=0$,(metric a covariant constant) and use the conventions $\eta^{\mu\nu}=diag(1,-1,-1,-1)$, for 
the metric and $\epsilon_{123}=1$, $\epsilon^{123}=-1$,for the Levi-Civita symbol. The mathematical structure of torsion 
theories has been elucidated in \cite{ Hehl}.

Coupling torsion to electromagnetism in a gauge-invariant fashion was first carried out by Novello \cite{Novello}, De Sabbata
and Gasperini \cite{Sabbata} and Duncan, Kaloper and Olive \cite{Duncan}. They pointed out that if the dual of the
torsion tensor, $T^\mu= \frac{1}{2}\epsilon^{\mu \nu \alpha \beta}T_{\nu \alpha \beta}$  is the divergence of a scalar,
$T^\mu =\partial^\mu \phi$, then it can be coupled to electromagnetic interactions in a gauge invariant way by the
interaction $T_\mu A_\nu \tilde F^{\mu \nu}= \phi F_{\mu \nu}\tilde F^{\mu \nu}$. A propagating scalar give rise to
long range forces which can be constrained by the observations of energy loss of Hulse-Taylor binary pulsar \cite{Mohanty}.
A cosmological scalar field which couples to electromagnetism causes birefringence of electromagnetic waves causing
polarization of radio signals from distant galaxies as pointed out by Harari and Sikivie \cite{Harari}. Carroll and Field
\cite{Carroll3} put strong constraints on the presence of cosmological background from the polarization of radio galaxies.
 Hammond \cite{Hammond} introduced an antisymmetric two-index torsion potential $\psi_{\mu \nu}$ , related
to the torsion tensor by $T_{\alpha \mu \nu} =\partial_{[\alpha} \psi_{\mu \nu ]}$ and coupled  it to electromagnetism through an
interaction of the form $F^{\mu \nu}\psi_{\mu \nu}$. This interaction is similar to the kinetic mixing between photons and
para-photons studied by Masso and Redondo \cite{Masso1, Masso2}.
If the scalar-torsion fields have heavier mass, then there are laboratory constraints on their coupling to electromagnetism
in optical polarization of lasers in external magnetic field \cite{ Cast, PVLAS} as proposed in \cite{Maiani, Raffelt} and in
long range non-Newtonian forces \cite{Dupays}.
Constraints on heavy propagating torsion which may be produced in accelerators or in supernovae have been surveyed in
\cite{Carroll3, Shapiro}.

In this paper we study non-minimal coupling of a cosmological background torsion field to electromagnetic field of the form
$\xi_{1} T^{\alpha\lambda}_{\quad \rho} \,F_{\alpha\nu} \,\partial_{\lambda}
\tilde{F}^{\rho\nu}$ and  $\xi_{2} T^{\sigma\gamma}_{\quad \delta} \,F_{\sigma\nu}\, \partial_{\gamma}
{F}^{\delta\nu}$. We find that the first type of coupling gives rise to birefringence
in the electromagnetic waves propagating through a background torsion field over cosmological distances.
The $\xi_{1}$ type of coupling is similar in structure  to the form $ \frac{1}{E_{P}} n^{\alpha} F_{\alpha \sigma} n \cdot \partial
(n_{\beta} \tilde F^{\beta \sigma}) $ proposed by Myers and Pospelov \cite{myers} motivated by quantum gravity arguments.
For second type of coupling, the transverse modes of electromagnetic waves propagates along null geodesics despite 
the presence of torsion coupling. Therefore, the  $\xi_{2}$ type of coupling has no effect on the wave propagation. 
These types of couplings are also similar in form to the phenomenological Lorentz violating couplings of electromagnetism studied
in Kostelecky and Mewes \cite{Kostelecky} who proposed that such couplings will mix the $E$ and $B$ mode CMB polarization due to
birefringence . A detailed study of the birefringence arising from Myers-Pospelov interaction on the CMB polarization was done
by Gubitosi et al. \cite{cooray}. By an analysis of the CMB polarization data from WMAP and BOOMERanG, Gubitosi et al. put a limit
on the angle of rotation of the plane of polarization of CMB signal at $\alpha =\left(-2.4 \pm 1.9\right)^{\circ}$.We use this result
of the angle of rotation of CMB to put the limits on the non-minimal coupling of torsion with electromagnetism.

%%%%%%%%%%%%%%%%%%%%%%%%%%%%%%%%%%%%%%%%%%%%%%%%%%%%%%%%%%%%%%%%%%%%%%%%%%%%%%%%%%%%%%%%%%%%%%%%%%%%%%
\section{Non-minimal torsion coupling with electrodynamics}
We start with the Lagrangian for the electromagnetic field on a manifold with torsion but no curvature. We
consider two non-minimal couplings of torsion with electromagnetism described by,
\begin{eqnarray}
{\cal L}=-\frac{1}{4} F_{\mu\nu}F^{\mu\nu} + \xi_{1} T^{\alpha\lambda}_{\quad \rho} \,F_{\alpha\nu}\left( \partial_{\lambda}
\tilde{F}^{\rho\nu}\right)+ \xi_{2} T^{\sigma\gamma}_{\quad \delta} \,F_{\sigma\nu}\left( \partial_{\gamma}
{F}^{\delta\nu}\right)
\end{eqnarray}
We will consider the two coupling terms separately.
%%%%%%%%%%%%%%%%%%%%%%%%%%%%%%%%%%%%%%%%%%%%%%%%%%%%%%%%%%%%%%%%%%%%%%%%%%%%%%%%%%%%%%%%%%%%%%%%%%%%%%%%%
\subsection{Case-I : when $\xi_{1} \neq 0$ and $\xi_{2}$ = 0}
Using the Euler-Lagrangian equation,
\begin{eqnarray}
 \partial_{\mu}\left(\frac{\partial \cal L}{\partial\left(\partial_{\mu}A_{\nu}\right)}\right)-\frac{\partial \cal L}{\partial A_{\nu}}=0
\end{eqnarray}
the equation of motion will be,
\begin{eqnarray}
\partial_{i}\,E^{i}& = &\xi_{1}\,\,T^{\alpha \lambda}_{\quad i}\,\,\partial_{\alpha}\,\partial_{\lambda}\,B^{i}\\
\label{eom1}
\partial_{0}\,E^{j}+\partial_{i}\,\epsilon^{0 ijk}B_{k}& = &\xi_{1}\,\,\partial_{\alpha}\,\partial_{\lambda}\,
\left[-T^{\alpha \lambda}_{\quad 0}B^{j}+T^{\alpha \lambda}_{\quad i}\epsilon^{0 ijk}E_{k}\right]
\end{eqnarray}
and from the Maxwell's equation for $F_{\mu \nu}$ ,
\begin{eqnarray}
\label{cond1}
\partial_{k}\,\epsilon^{0 kjl}E_{l}-\partial_{0}B^{j}&=&0\\
\partial_{j}B^{j}&=&0
\end{eqnarray}
Using the eq.~(\ref{cond1}),the eq.~(\ref{eom1}) can be written as,
\begin{eqnarray}
-\partial^{0}\partial_{0}E^{j}-\partial^{m}\left(\partial_{m}E^{j}-\partial^{j}E_{m}\right)=-\xi_{1}\,\,\partial_{\alpha}\,\partial_{\lambda}\,
\left[T^{\alpha \lambda}_{\quad 0}\,\epsilon^{0 mjn}\,\partial_{m}E_{n}-T^{\alpha \lambda}_{\quad i}\,\epsilon^{0 ijk}\,\partial^{0}E_{k}\right]
\label{eom2}
\end{eqnarray}
which is the equation of motion for the electric field.
We consider the plane wave solution where, the electric field can be written as,
$E_{m}(x)=E_{m}(k)\,e^{-i k^{\rho}x_{\rho}}$ and eq.~(\ref{eom2}) reduces to,
\begin{eqnarray}
k^{0}k_{0}E^{j}+k^{m}\left(k_{m}E^{j}-k^{j}E_{m}\right)=-i\,\xi_{1}\,\,k_{\alpha}\,k_{\lambda}\,
\left[T^{\alpha \lambda}_{\quad 0}\epsilon^{0 mjn} k_{m}E_{n}+T^{\alpha \lambda}_{\quad i}\epsilon^{0 ijk} k^{0}E_{k}\right]
\label{eom3}
\end{eqnarray}
Choosing Z-axis along the direction of propagation i.e. $k_{\alpha}=(\omega,0,0,p)$,
the three equations of motion given by the eq.~(\ref{eom3}) take the form,
\begin{eqnarray}
\left(\omega^{2}-p^{2}\right)E^{1}&=&2i\,\xi_{1}\, T_{1}\, p^{3}\, E^{2}\nonumber\\
\left(\omega^{2}-p^{2}\right)E^{2}&=&-2i\,\xi_{1}\, T_{1}\, p^{3}\, E^{1}\nonumber\\
\omega^{2}E^{3}&=& -i\, \xi_{1}\, \left(T_{2}\, E^{1}-T_{3}\, E^{2} \right)p^{3}
\end{eqnarray}
where,
\begin{eqnarray}
T_{1}&=&T^{0 0}_{\quad 3}+T^{3 3}_{\quad 0}\,\,,\nonumber\\
T_{2}&=&T^{0 0}_{\quad 2}+T^{0 3}_{\quad 2}+T^{3 0}_{\quad 2}+T^{3 3}_{\quad 2}\,\,,\nonumber\\
T_{3}&=&T^{0 0}_{\quad 1}+T^{0 3}_{\quad 1}+T^{3 0}_{\quad 1}+T^{3 3}_{\quad 1}\nonumber\,\,.
\end{eqnarray}
The equation of motion for the transverse modes are,
\begin{eqnarray}
\left(
  \begin{array}{ccc}
    \left(\omega^2-p^2\right)  & &  2\,i\,\xi_{1}\,T_{1}\,p^{3} \\
   - 2\,i\,\xi_{1}\,T_{1}\,p^{3} & &  \left(\omega^2-p^2\right)  \\
  \end{array}\right)\left(
               \begin{array}{c}
                 E^{1} \\
                 E^{2} \\
               \end{array}
             \right)=0
\label{d.rel}
\end{eqnarray}
Therefore, the equations of motion for the right- and left- circularly polarized fields $E_{\pm}=E^{1}{\hat e}_{x}\pm i\,E^{2}{\hat e}_{y}$
turn out to be,
\begin{eqnarray}
\left(\omega^{2}-p^{2}\right)\left(E^{1}\pm iE^{2}\right) \mp 2\,\xi_{1}\,p^{3}\,T_{1}\left(E^{1}\pm iE^{2}\right)=0
\end{eqnarray}
From the dispersion relation~(\ref{d.rel}), $\omega_{\pm}$ will be;
\begin{eqnarray}
\omega_{\pm}=p\left(1\pm \xi_{1}\,p\,T_{1}\right)\nonumber\\
\end{eqnarray}
The rotation of the plane of polarization can be written in terms of the difference of
the $\omega_{\pm}$ as following,
\begin{eqnarray}
\alpha&=&\left(\omega_{-}-\omega_{+}\right)t\nonumber\\
&=&2\,\xi_{1}\,\,p^{2}\,\,T_{1}\,\,t
\label{alpha}
\end{eqnarray}
where $ t $ is the propagation time. We notice that only the combination, $T_{1}=T^{0 3}_{\quad 0}-T^{3 3}_{\quad 0}$
enters the equation because of our choice of coordinates. Experimental constraints on $\alpha$ have been put
$\left(-2.4 \pm 1.9\right)^{\circ}$ from the observation of CMB polarization by WMAP and BOOMERanG. We will make use
of these constraints to put bounds on the torsion coupling $\xi_{1}T_{1}$.
%%%%%%%%%%%%%%%%%%%%%%%%%%%%%%%%%%%%%%%%%%%%%%%%%%%%%%%%%%%%%%%%%%%%%%%%%%%%%%%%%%%%%%%%%%%%%%%%%%%%%%%%%%%%%%%%%%%
\subsection{Case-II :  when $\xi_{2} \neq 0$ and $\xi_{1}$ = 0}
Putting $\xi_{1}=0$ in the correction term in the lagrangian, the equations of motion will be,
\begin{eqnarray}
\partial_{i}\,E^{i}& = &\xi_{2}\,\,T^{\alpha \lambda}_{\quad i}\,\,\partial_{\alpha}\,\partial_{\lambda}\,E^{i}\\
\label{2.eom1}
-\partial_{0}\,E^{j}+\partial_{i}\,\epsilon^{0 ijk}B_{k}& = &\xi_{2}\,\,\partial_{\alpha}\,\partial_{\lambda}\,
\left[-T^{\alpha \lambda}_{\quad 0}E^{j}+T^{\alpha \lambda}_{\quad i}\epsilon^{0 ijk}B_{k}\right]
\end{eqnarray}
Using again the Maxwell's eq.~(\ref{cond1}), eq.~(\ref{2.eom1}) can be written as,
\begin{eqnarray}
-\partial^{0}\partial_{0}E^{j}-\partial^{m}\left(\partial_{m}E^{j}-\partial^{j}E_{m}\right)=-\xi_{2}\,\,\partial_{\alpha}\,\partial_{\lambda}\,
\left[T^{\alpha \lambda}_{\quad 0}\partial^{0}E^{j}-T^{\alpha \lambda}_{\quad i}\left(\partial^{j}E^{i}-\partial^{i}E^{j}\right)\right]
\label{2.eom2}
\end{eqnarray}
which is the equation of motion for the electric field.
Considering again the plane wave solution where, the electric field can be written as,
$E_{m}(x)=E_{m}(k)\,e^{-i k^{\rho}x_{\rho}}$, eq.~(\ref{2.eom2}) reduces to,
\begin{eqnarray}
k^{0}k_{0}E^{j}+k^{m}\left(k_{m}E^{j}-k^{j}E_{m}\right)=-i\,\xi_{2}\,\,k_{\alpha}\,k_{\lambda}\,
\left[T^{\alpha \lambda}_{\quad 0} k^{0}E^{j}-T^{\alpha \lambda}_{\quad i}\left(k^{j}E^{i}-k^{i}E^{j}\right)\right]
\label{2.eom3}
\end{eqnarray}
With the choice of Z-axis along the direction of propagation i.e. $k_{\alpha}=(\omega,0,0,p)$,
the three equations of motion given by the eq.~(\ref{eom3}) take the form,
\begin{eqnarray}
\left(\omega^{2}-p^{2}\right)E^{1}&=&0\nonumber\\
\left(\omega^{2}-p^{2}\right)E^{2}&=&0\nonumber\\
\omega^{2}E^{3}&=&- i\, \xi_{2}\, \left[T_{1}\,E^{3}+\left(T_{3}\, E^{1}+T_{2}\, E^{2} \right)\right]p^{3}
\end{eqnarray}
Here, we can see that the transverse modes propagates along null geodesics despite the presence of torsion coupling. 
So, we conclude that a coupling of the form $\xi_{2} T^{\sigma\gamma}_{\quad \delta} \,F_{\sigma\nu}\left( \partial_{\gamma}
{F}^{\delta\nu}\right)$ has no effect on the wave propagation. Therefore, we can not put the limits on $\xi_{2}$ from the
consideration of CMB polarization.
%%%%%%%%%%%%%%%%%%%%%%%%%%%%%%%%%%%%%%%%%%%%%%%%%%%%%%%%%%%%%%%%%%%%%%%%%%%%%%%%%%%%%%%%%%%%%%%%%%%
\section{Polarization of Cosmic Microwave Background}
Thomson scattering of CMB photons on the last scattering surface gives rise to linear polarization.
This linear polarization can be expressed by the two Stokes parameters Q and U.
The Boltzmann equation, which describes the time evolution of the polarization perturbation will be,
\begin{eqnarray}
\dot{\Delta}_Q+ik\mu\Delta_Q&=&-\dot{\tau}\left[\Delta_Q+\frac{1}{2}\left(1-P_{2}(\mu)
\right)S_{p}\right]
\label{Q}\\
\dot{\Delta}_U+ik\mu\Delta_U&=&-\dot{\tau}\Delta_U
\label{U}
\end{eqnarray}
where
\begin{eqnarray}
S_{p}&=&\Delta_{T_{2}}+\Delta_{Q_{2}}-\Delta_{Q_{0}}\nonumber\,,\\
\dot{\tau}&=&\frac{x_{e}n_{e}\sigma_{T}a}{a_{0}}\nonumber\,.
\end{eqnarray}
and $x_{e}$ the ionization fraction, $n_{e}$ the electron number density, $\sigma_{T}$ the
Thomson scattering cross section, and a is the scale factor.
If any physical mechanism, which causes the rotation of the polarization plane, then the
time evolution of the stokes parameters will be modified.
Considering only the scalar perturbation, the modified Boltzmann equation will be,
\begin{eqnarray}
\dot{\Delta}_Q+ik\mu\Delta_Q&=&-\dot{\tau}\left[\Delta_Q+\frac{1}{2}\left(1-P_{2}(\mu)
\right)S_{p}\right]+2w\Delta_{U}
\label{Q0}\\
\dot{\Delta}_U+ik\mu\Delta_U&=&-\dot{\tau}\Delta_U-2w\Delta_{Q}
\label{U0}
\end{eqnarray}
 and $\omega$ is the rate of change of angle of the polarization
due to new physics, which in this case is the background torsion.\\
Changing the variable as \cite{kosowsky} ,
\begin{eqnarray}
\tilde{\Delta}_{Q}&\equiv&e^{ik\mu\eta-\tau}\Delta_{Q}\nonumber\,,\\
\tilde{\Delta}_{U}&\equiv&e^{ik\mu\eta-\tau}\Delta_{U}\nonumber\,,\\
\tilde{S_{p}}&\equiv&e^{ik\mu\eta-\tau}S_{p}\nonumber\,.\\
\end{eqnarray}
Eq.~(\ref{Q0}) and eq.~(\ref{U0}) reduce to the form,
\begin{eqnarray}
\dot{\tilde{\Delta}}_Q&=&-\dot{\tau}\frac{1}{2}\left(1-P_{2}(\mu)\right)\tilde{S}_{p}+2w\tilde{\Delta}_{U}
\label{Q1}\\
\dot{\tilde{\Delta}}_U&=&-2w\tilde{\Delta}_{Q}
\label{U1}
\end{eqnarray}
These two eqs.~(\ref{Q1}) and eqs.~(\ref{U1}) can be combined as,
\begin{eqnarray}
\dot{\tilde{\Delta}}_Q\pm i\dot{\tilde{\Delta}}_U=-\dot{\tau}\frac{1}{2}\left(1-P_{2}\left(\mu\right)\right)
\tilde{S}_{p}\mp2wi\left(\tilde{\Delta}_{Q}\pm i\tilde{\Delta}_{U}\right)
\label{Q+iU}
\end{eqnarray}
Now define,
\begin{eqnarray}
\tilde{\Delta}_{Q}\pm i\tilde{\Delta}_{U}=P(\eta)
\end{eqnarray}
Eqs.~(\ref{Q+iU}) can be written as,
\begin{eqnarray}
\dot{P}(\eta)\pm2\omega iP(\eta)&=&-\dot{\tau}\frac{1}{2}\left(1-P_{2}\left(\mu\right)\right)\tilde{S}_{p}\nonumber\\
&=&-\frac{3}{4}\dot{\tau}(\eta)\left(1-\mu^{2}\right)\tilde{S}_{p}
\label{P.Eq1}
\end{eqnarray}
Eq.~(\ref{P.Eq1}) can be solved for $P(\eta)$ to give
\begin{eqnarray}
P(\eta)=\left[-\frac{3}{4}\left(1-\mu^{2}\right)\int\dot{\tau}(\eta)\tilde{S}_{p}d\eta
e^{\pm2i\int\omega(\eta)d\eta}\right]e^{\mp2i\int\omega(\eta)d\eta}
\label{P_eq}
\end{eqnarray}
Therefore,$\Delta_{Q}$ and $\Delta_{U}$ will be,
\begin{eqnarray}
\Delta_{Q}&=&\left[-\frac{3}{4}\left(1-\mu^{2}\right)\left(\int\dot{\tau}(\eta)\tilde{S}_{p}
e^{\pm2i\int\omega(\eta)d\eta}\,d\eta\right)\cdot e^{ik\mu\eta-\tau}\right]\cos\left(2\alpha\right)\\
\Delta_{U}&=&-\left[-\frac{3}{4}\left(1-\mu^{2}\right)\left(\int\dot{\tau}(\eta)\tilde{S}_{p}
e^{\pm2i\int\omega(\eta)d\eta}\,d\eta\right) \cdot e^{ik\mu\eta-\tau}\right]\sin\left(2\alpha\right)
\label{result}
\end{eqnarray}
where the angle of rotation $\alpha=\int\omega(\eta)d\eta$.

 When $\alpha=0$ i.e. $\omega=0$, then
\begin{eqnarray}
\Delta_{Q}|_{\omega=0}&=&\left[-\frac{3}{4}\left(1-\mu^{2}\right)\left(\int\dot{\tau}(\eta)\tilde{S}_{p}
e^{\pm2i\int\omega(\eta)d\eta}\,d\eta\right)\cdot e^{ik\mu\eta-\tau}\right]\nonumber\\
\Delta_{U}|_{\omega=0}&=&0\nonumber
\end{eqnarray}
If both scalar and tensor perturbation are present,there will be non zero $\Delta_{U}$, even when $\alpha=0$.
\begin{eqnarray}
\Delta_{Q}&=&\Delta_{Q}|_{\omega=0}\cos\left(2\alpha \right)+\Delta_{U}|_{\omega=0}\sin\left(2\alpha \right)\\
\Delta_{U}&=&-\Delta_{Q}|_{\omega=0}\sin\left(2\alpha \right)+\Delta_{U}|_{\omega=0}\cos\left(2\alpha \right)
%\label{result}
\end{eqnarray}
In general, these stokes parameter Q and U can be related to E and B modes of polarization as follows in \cite{Seljak},
\begin{eqnarray}
\left(Q \pm i U\right)\left(\eta\,,\,\overrightarrow{x}\,,\,\hat{\eta}\right)=
\int \frac{d^{3}\textbf{q}}{(2\pi)^{3}}\, \sum \, \sum_{m=-2}^{2} \, \, \left(E_{l}^{(m)} \pm B_{l}^{(m)}\right)\, _{\pm 2} G_{l}^{m}
\label{Q=EandU=B}
\end{eqnarray}
where, $ _{s} G_{l}^{m}(\overrightarrow x,\hat n)=(-i)^{l}\sqrt{\frac{4\pi}{2l+1}}\,\left[\,_{s}\, Y_{l}^{m}(\hat n)\right]\, exp(i \overrightarrow k \cdot \overrightarrow x)$.

Using the expression for the power spectra $C_{l}^{XY} \sim \int dk \left[k^{2} \Delta_{X}\Delta_{X}\right]$ , $X,Y=T,E,B$ ;
we can derive the correlations for T, E and B for CMB  in terms of $\alpha$ as follows,
\begin{eqnarray}
C_{l}^{EE}&=&\tilde{C_{l}}^{EE}\cos^{2}(2\alpha)+\tilde{C_{l}}^{BB}\sin^{2}(2\alpha)\,\,,\nonumber\\
C_{l}^{BB}&=&\tilde{C_{l}}^{EE}\sin^{2}(2\alpha)+\tilde{C_{l}}^{BB}\cos^{2}(2\alpha)\,\,,\nonumber\\
C_{l}^{EB}&=&\frac{1}{2}\left(\tilde{C_{l}}^{EE}-\tilde{C_{l}}^{BB}\right)\sin(4\alpha)\,\,,\nonumber\\
C_{l}^{TE}&=&\tilde{C_{l}}^{TE}\cos(2\alpha)\,\,,\nonumber\\
C_{l}^{TB}&=&\tilde{C_{l}}^{TE}\sin(2\alpha)\,\,.
\end{eqnarray}

From these relations we see that an initial E-mode polarization generated by Thomson scattering in the last scattering surface 
can source a B-mode polarization during propagation through a torsion background. 

Using the polarization data from WMAP and BOOMERanG, Gubitosi et al. \cite{cooray} have put the limits on the rotation of the
plane of polarization $\alpha =\left(-2.4 \pm 1.9\right)^{\circ}$. Using the relation~(\ref{alpha}), the values
$H_{\circ}=72\left(Km/s\right)/Mpc$, $z=1100$, $E_{p}=1.22\times 10^{19}$, $p=100GHz$, $\Omega_{m}=0.3$, $\Omega_{\lambda}=0.7$
and taking the time of propagation as,\begin{eqnarray}
t=\frac{1}{H_{\circ}}\int^{Z}_{0}\frac{\left( 1+z\right)}{\sqrt{\Omega_{m}(1+z)^{3}+\Omega_{\Lambda}}}dz \nonumber
\end{eqnarray}
we can convert this to limits on the background torsion as,
\begin{eqnarray}
\xi_{1}T_{1}=\left(-3.35 \pm 2.65\right)\,\times\,10^{-22}\,\,\,GeV^{-1}
\end{eqnarray}
The limits on $\xi_{1}T_{1}$ is comparable to the limits on the Lorentz violating interaction of electromagnetism studied in \cite{Kostelecky}.
%%%%%%%%%%%%%%%%%%%%%%%%%%%%%%%%%%%%%%%%%%%%%%%%%%%%%%%%%%%%%%%%%%%%%%%%%%%%%%%%%%%%%%%%%%%%%%%%%%%%%%%%%%%%%%%%%%%%%
\section{Discussions}
We have studied non-minimal coupling of electromagnetism to torsion background of the form
$\xi_{1} T^{\alpha\lambda}_{\quad \rho} \,F_{\alpha\nu}\left( \partial_{\lambda} \tilde{F}^{\rho\nu}\right)$
and $\xi_{2} T^{\sigma\gamma}_{\quad \delta} \,F_{\sigma\nu}\left( \partial_{\gamma} {F}^{\delta\nu}\right)$.
The first of these interaction is similar in structure to the non-minimal coupling 
$ \frac{1}{E_{P}} n^{\alpha} F_{\alpha \sigma} n \cdot \partial (n_{\beta} \tilde F^{\beta \sigma}) $,
studied by Myers-Pospelov \cite{myers}, where $n^{\alpha}$ is some fixed lorentz violating vector.In our case,
the role of $n^{\alpha}$ is played by the background torsion.The second interaction
$\xi_{2} T^{\sigma\gamma}_{\quad \delta} \,F_{\sigma\nu}\left( \partial_{\gamma} {F}^{\delta\nu}\right)$ has the interesting
property that irrespective of the magnitude of $\xi_{2} T$, the electromagnetic waves propagate along null geodesics
and suffer no polarization and  one can't put the constraints on this coupling from the observation of the CMB.
%%%%%%%%%%%%%%%%%%%%%%%%%%%%%%%%%%%%%%%%%%%%%%%%%%%%%%%%%%%%%%%%%%%%%%%%%%%%%%%%%%%%%%%%%%%%%%%%%%%%%%%%%%%%%%%%%%%%%
%\newpage

\end{document}